\documentclass[acmlarge,screen]{acmart}

\AtBeginDocument{%
  \providecommand\BibTeX{{%
    \normalfont B\kern-0.5em{\scshape i\kern-0.25em b}\kern-0.8em\TeX}}}

\setcopyright{acmcopyright}
\copyrightyear{2019}
\acmYear{2019}
\acmDOI{10.1145/1122445.1122456}

\usepackage[linesnumbered,ruled,vlined]{algorithm2e}
\usepackage{booktabs}

\begin{document}

\title{Design Methodology for Energy Efficient Unmanned Aerial Vehicles}

\author{Jingyu He}
\affiliation{%
  \institution{University of Southern California}
  \streetaddress{3740 McClintock Ave}
  \city{Los Angeles}
  \state{CA}
  \country{United States}}
\email{jingyuhe@usc.edu}

\author{Yao Xiao}
\affiliation{%
  \institution{University of Southern California}
  \streetaddress{3740 McClintock Ave}
  \city{Los Angeles}
  \state{CA}
  \country{United States}}
\email{xiaoyao@usc.edu}

\author{Corina Bogdan}
\affiliation{%
  \institution{Northeastern University}
  \streetaddress{360 Huntington Ave}
  \city{Boston}
  \state{MA}
  \country{United States}}
\email{i.bogdan@northeastern.edu}

\author{Shahin Nazarian}
\affiliation{%
  \institution{University of Southern California}
  \streetaddress{3740 McClintock Ave}
  \city{Los Angeles}
  \state{CA}
  \country{United States}}
\email{shahin.nazarian@usc.edu}

\author{Bogdan Paul}
\affiliation{
  \institution{University of Southern California}
  \streetaddress{3740 McClintock Ave}
  \city{Los Angeles}
  \state{CA}
  \country{United States}}
\email{pbogdan@usc.edu}

\renewcommand{\shortauthors}{He, et al.}

\begin{abstract}
In this paper, we present a load-balancing approach to analyze and partition the UAV perception and navigation intelligence (PNI) code for parallel execution, as well as assigning each parallel computational task to a processing element in an Network-on-chip (NoC) architecture such that the total communication energy is minimized and congestion is reduced. First, we construct a data dependency graph (DDG) by converting the PNI high level program into Low Level Virtual Machine (LLVM) Intermediate Representation (IR). Second, we propose a scheduling algorithm to partition the PNI application into clusters such that (1) inter-cluster communication is minimized, (2) NoC energy is reduced and (3) the workloads of different cores are balanced for maximum parallel execution. Finally, an energy-aware mapping scheme is adopted to assign clusters onto tile-based NoCs. We validate this approach with a drone self-navigation application and the experimental results show that our optimal 32-core design achieves an average 82\% energy savings and 4.7x performance speedup against the state-of-art flight controller.
\end{abstract}

%
%



\maketitle

\section{Introduction}
Unmanned aerial vehicles (UAVs) are emerging as critical tools for mapping large areas, patrolling, searching, and rescuing applications. These tasks are usually dangerous, repetitive and have to be carried out in extreme conditions, making them ideal for autonomous drones. From the traditional quadrotors \cite{hoffmann2007quadrotor} to a fully-actuated hexarotors \cite{rajappa2015modeling}, and to the futuristic volocopoters \cite{ullman2017comparing}, many researchers demonstrated various designs of the UAVs for different application scenarios, as shown in Fig. 1. 
The increasingly complicated self-navigation and collision-avoiding applications of the evolving UAVs ask for a high-performance and low-power flight controller. 

\begin{figure}[h]
  \centering
  \includegraphics[scale=0.4]{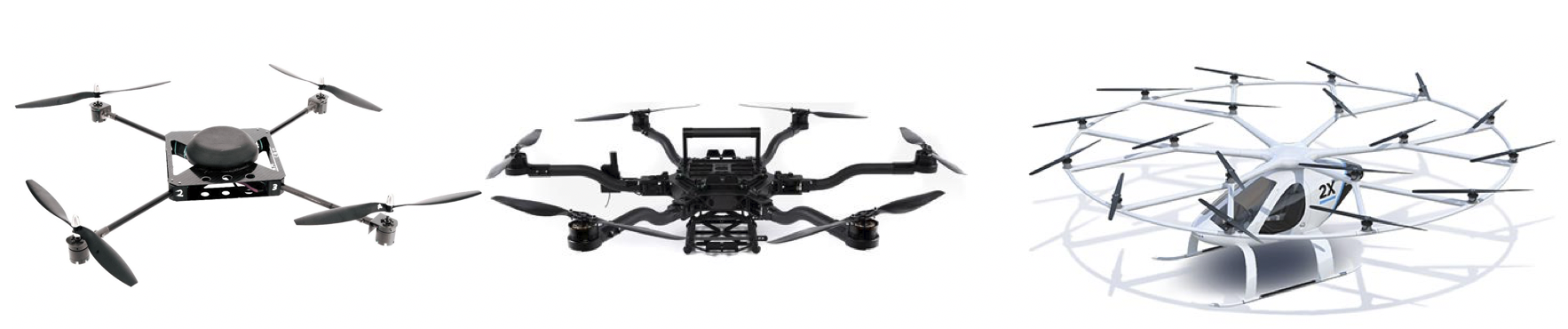}
  \caption{Various multirotos: quadrotor (left), hexarotor (middle), volocoptor (right)}
\end{figure}

We cannot stress the importance of the performance of flight control applications enough. The lack of data-processing speed of the flight control computer chips has led to plane crashes reportedly. At the same time, low-power design is critical for UAVs as well. One reason is that high power dissipation brings tremendous cooling challenges to maintain the hardware at a suitable temperature. Another is that batteries are the only energy source for drones, limiting the running time of drones. 

In order to push the performance and energy boundary of systems-on-chips, Dally and Towles \cite{dally2001route} proposed the tile-based Network-on-chips (NoC) as the ideal architecture for scalable and low-power on-chip communication. Such chips use tiles as building blocks such as CPUs, GPUs, ASIC and memory. A standard interface is embedded into each tile to route flits for communication. There have been many previous studies on energy-aware NoC designs. In contrast to prior NoC work, the goal of this paper is to investigate the parallelization of the UAV perception and navigation intelligence while taking the computation and communication power consumption into consideration. As shown in Fig. 2, we first compile the navigation program into LLVM IR and construct the DDG, where each node denotes only a useful instruction with its power consumption and each edge represents the data dependency with the weight being data size times latency. Second, based on DDG graph, we propose a scheduling algorithm to partition the PNI application into clusters such that (1) inter-cluster communication is minimized, (2) NoC energy is reduced and (3) the workloads of different cores are balanced for maximum parallel execution. Finally, we incorporate topological sort into the our energy-aware mapping scheme to further reduce static power consumption resulted by congestion.

Towards this end, the main contributions of this paper are as follow:
\begin{itemize}
	\item To the best of our knowledge, our work is the first to incorporate the static energy consumption analysis of application into a compiler-based task partition.
	\item Besides volume, we propose a mapping strategy to also consider the timing of inter-core communications, reducing the congestion time and static energy consumption of hardware resources.
\end{itemize}
The rest of the paper is organized as follows: Section II discusses the related work. Section III introduces the basics of UAV control. Section IV illustrates the energy model for NoCs, the load-balancing and energy-aware community detection algorithm, and the low-power mapping. Section V validates the framework and shows experimental results compared to the baseline model.

\begin{figure}[h]
  \centering
  \includegraphics[width=\linewidth]{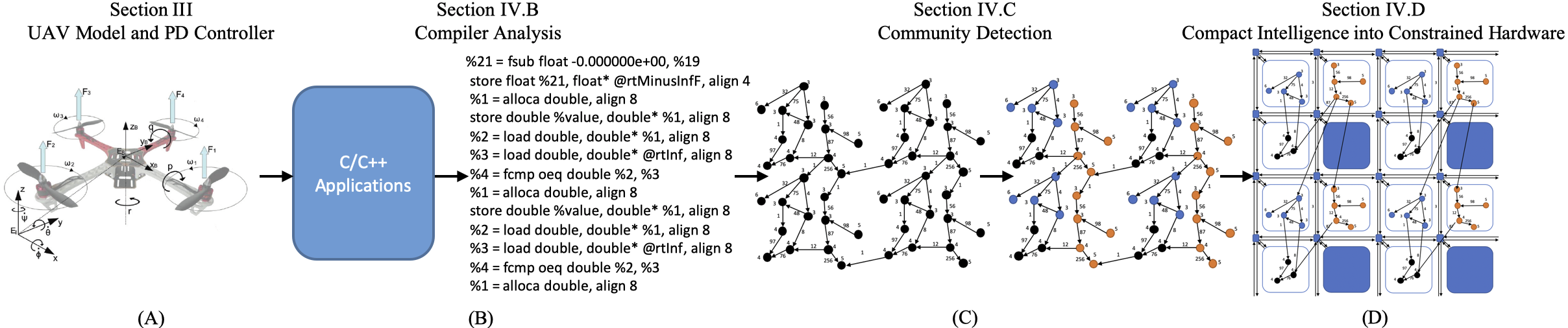}
  \caption{Overview of the UAV intelligent processing architecture workflow. (A): An UAV and its control basics. (B): The perception and navigation intelligence application (as a high level program) is compiled into LLVM IR trace through compiler analysis. This allows to remove the unnecessary computation and communication overhead of high level programs. (C): We transform the trace into the DDG and detect communities. (D): Each processing community is mapped onto an NoC processing element in such a way that its communication energy is minimized and congestion is reduced. The unused cores are clock-gated to save energy, indicated by the blue tiles.}
\end{figure}

\section{Related Work}
As UAV's core avionic part, flight controller is a system that integrates hardware and software to make drones fully capable of flying without human beings' guidance. Basic linear controllers such as PID or LQR have been widely used \cite{bouabdallah2004pid, cowling2010direct} due to their simplicity in design and implementation. However, they are mostly limited to simple non-agile use cases. To model the non-linearity of the motion control more effectively, nonlinear controllers such as backstepping and feedback linearization were proposed \cite{lee2009feedback}. In addition, learning-based controllers were proposed for acrobatic moves \cite{lupashin2010simple}. They utilize the approximation ability of the neural networks for learning the dynamics of the multirotors. 

To address the increasing computation and energy demand of flight controllers, there has been a significant amount of previous research on energy-aware and load-balancing scheduling and mapping on multicore embedded systems. From a mathematical and control perspective, Bogdan et al. in \cite{bogdan2015mathematical,bogdan2015mathematical1} provide a complex approach to dynamically characterize the workload of multicore systems for performance and power optimization. Xiao et al. propose a complex network-inspired application partitioning tool to improve multicore parallelization \cite{xiao2017load}. Tan et al. develop a low-power customizable manycore architecture for wearables using a lightweight message-passing scheme \cite{tan2018locus}. Navion \cite{suleiman2019navion} design an energy-efficient accelerator to fully integrate visual-inertial odometry system-on-chip while eliminating expansive off-chip processing and storage for autonomous navigation of drones. In terms of mapping and routing, an efficient branch-and-bound algorithm proposed by Hu et al. \cite{hu2003exploiting} automatically maps the IPs onto a generic NoC so that the communication cost is minimized while the timing constraint is met. Lyons et al. create the accelerator store by sharing memories between accelerators to combine the scalability of homogeneous multi-core and SoC’s high performance and power-efficient hardware accelerators \cite{lyons2012accelerator}. Esmaeilzadeh et al.'s work \cite{esmaeilzadeh2011dark} provides a novel and extendible model that integrates device scaling trends, core design tradeoffs, and multicore configurations. In contrast to prior work, we present an energy-aware load-balancing community detection algorithm together with a mapping strategy and test it using a UAV self-navigation application.

\section{Brief Overview of the Basics of the UAV Navigation Controller}

Fig. 2(A) shows a UAV with six degrees of freedom. Three degrees of freedom describe the translational motions ($x, y, z$) and the other three are the rotational motions ($r,p,q$). Each of the four propellers is equipped with a rotor providing the angular velocity. These four angular velocities correspond to the inputs of the quadrotor, $\omega_{i}=[\omega_{1},\omega_{2},\omega_{3},\omega_{4}]$. Twelve outputs are generated from the quadrotor, $X=[x,y,z,r,p,q,\dot{x},\dot{y},\dot{z},\dot{r},\dot{p},\dot{q}]$, corresponding to the translational and rotational positions, and their corresponding velocities \cite{corke2017flying}.

For real-time applications, the error between the actual UAV position, estimated by a navigation system, and the desired position is fed into a PD-controller to determine the required control inputs. The required rotor speeds are then calculated from the respective torques using:
\begin{equation}
 \begin{pmatrix}T\\ \Gamma \end{pmatrix} = \begin{pmatrix}-b&-b&-b&-b\\0&-db&0&db\\-db&0&db&0\\k&-k&k&-k\end{pmatrix}  \begin{pmatrix}{\omega_1^2}\\{\omega_2^2}\\{\omega_3^2}\\{\omega_4^2}\end{pmatrix}\label{eq}
\end{equation}
where $T$ is the thrust vector for each propeller, $\Gamma$ is the torque vector applied to the airframe, $b$ represents the lift constant, $d$ is the distance from the rotor to the center of the mass and $k$ is secondary lift constant. The control structure employed to fly the quadrotor can be found in \cite{corke2017flying,armah2016feedback}, and is based on Proportional Derivative action to get the quadrotor's attitude (roll, pitch, yaw) and altitude.

\section{Parallelization Discovery and Energy Optimization Approach}
\subsection{Energy Model}
Both IP cores and interconnection consume energy. While most of the mapping algorithms based on the one in \cite{hu2003exploiting} only compute dynamic energy, our model considers both static and dynamic power dissipation. N. Grech et al. \cite{grech2015static} propose an application static energy analysis technique to determine the instruction energy model directly at the LLVM IR level. Through analysis and measurement of a large set of target ISA instructions, it was found that LLVM IR instructions can be divided roughly into four groups: memory, $M$, program flow, $B$, division, $D$, and all other instructions, $G$. This yields an energy model $E_{N}$ of a program executed sequentially in a computing node:
\begin{equation}
E_{N} =  \sum_{i \in \lbrace M, B, D, G \rbrace}^{n}{E_{i}N_{i}}\label{eq}
\end{equation}
where $E_{i}$ is the energy cost of a single instruction in group $i$, $N_{i}$ is the number of the instructions executed in that group, and $n$ denotes the number of instructions.

Using the bit energy concept proposed by Ye et al. in \cite{ye2002analysis}, the total dynamic energy consumption can be computed by:
\begin{equation}
E_{DyNoC}=  \sum_{i=1}^{a}\sum_{j=1}^{b}{w_{ij}(\eta_{ij}E_{S_{bit}} + (\eta_{ij}  - 1)  \times  E_{L_{bit}})}\label{eq}
\end{equation}
where $E_{S_{bit}}$ and $E_{L_{bit}}$ represent the energy consumed by switch and link; $\eta_{ij}$ is the number of routers the packet from tile $\tau_{i}$ to tile $\tau_{j}$ passes through along the way; $w_{ij}$ is the size of the packet; $a$ and $b$ denote the number of tiles on $x$ and $y$ respectively. 

The static power is defined to characterize the energy consumed when packets are congested in the buffers. For simplicity, static power is defined as:
\begin{equation}
 E_{StNoC}=  \sum_{i=1}^{n}{P_{St} \times w_{i} \times t_{i}}\label{eq}
\end{equation}
where $n$ is the number of times that congestion occurs; $P_{St}$ is the energy consumption of one bit of data stored in the buffer for one unit of time; $w_{i}$ is the data size of the $i$th congestion; and $t_{i}$ is time of the $i$th congestion. Equation (5) gives the total energy consumption for the interconnect.
\begin{equation}
E_{NoC} = E_{StNoC} + E_{DyNoC}\label{eq}
\end{equation}

Finally, given the total number of tiles $n$, the energy consumption of the entire chip is computed as:
\begin{equation}
E = \sum_{i=1}^{n}E_{N_{i}} + E_{NoC}\label{eq}
\end{equation}

\subsection{Compiler Analysis and Model of Computation Extraction}
In order to generate the data dependency graph (DDG), we adopt the LLVM IR \cite{lattner2004llvm}. The rationale behind this is that LLVM is a language-independent system that exposes the commonly-used primitives to implement high-level language features, which makes it very easy to generate back-end for any target platform. 

With the help of Clang, C/C++ applications are compiled into a dynamic IR execution trace. We developed a parser to construct a data dependency graph from the IR trace. The parser analyzes memory operations to obtain latency and data sizes. Because the execution times and energy vary on data sizes and where the data resides, taking those values into account could potentially reduce inter-core communications by grouping the source and destination instructions of a register into one cluster. Three hash tables are created and updated when parsing: the source table, the destination table and the dependency table. The source/destination tables are used to keep track of source/destination registers with keys being source or destination registers and values being the corresponding line number. The dependency table is to store dependencies between nodes with keys being the line number for current instruction, and values being clock cycles, data sizes and line numbers of previous instructions dependent on the same virtual register.

\begin{table}
  \caption{The source, destination and weight tables}
  \label{tab:freq}
  \begin{tabular}{|c|c|c|c|c|c|}
    \toprule
    \multicolumn{6}{|c|}{LLVM IR trace}\\
    \midrule
    \multicolumn{6}{|p{7cm}|}{
    store double \%5, double* \%1, align 8 
    \newline \%2 = load double, double* \%1,     align 8 
    \newline \%3 = load double, double* \%6,     align 8 
    \newline \%4 = fcmp oeq double \%2, \%3 
    }\\ 
    \midrule
    \multicolumn{2}{|c|}{Src Table} & \multicolumn{2}{c|}{Dest Table} & \multicolumn{2}{c|}{Dependency Table}\\
    \midrule
    Key & Value & Key & Value & Key & Value\\
    \midrule
    \%5 & 1 & \%1 & 1 & 2 & 1\\
    \midrule
    \%1 & 2 &  \%2 & 2 & 4 & 2,3 \\
    \midrule
    \%6 & 3 &  \%3 & 3 & & \\
    \midrule
    \%2, \%3 &  4 & \%4 & 4 &  & \\
  \bottomrule
\end{tabular}
\end{table}

For example, in Table I, a LLVM IR snippet is extracted from an application compiled by Clang front-end. As the parser reads the first line, a source table and a destination table are created. The source table is updated with the key being \%5 and the value being 1 and its destination register is hashed into the destination table with the key being \%1 and value the being 1. When line 2 is read, the source register \%1 happens to be the destination register in line 1. A dependency table is created and updated with the key being 2 (line number of current instruction) and value being 1 (line number of the dependent instruction). Following the same procedure, the three hash tables will look like what is shown in Table 1.

\subsection{Discovering the Processing Community Structure}
To formulate this problem, we introduce the following concepts:
\begin{definition}
A data dependency graph (DDG) is a weighted directed graph $G = G(a_{i},b_{ij},e_{i},w_{ij}|i,j\in N|)$ where each vertex $a_{i}$ represents one LLVM IR instruction; each edge $b_{ij}$ with weights $w_{ij}$ characterizes either the dependency from $a_{i}$ to $a_{j}$ or the control flow such as jumps or branches from one block to another; and $e_{i}$ stands for the estimated energy of the vertex given in Section IV.A.
\end{definition}

\begin{definition}
A weight $w_{ij}$ between $a_{i}$ and $a_{j}$ is calculated by latency times data size. Latency characterizes the delay from $a_{i}$ to $a_{j}$ based on the timing information. Data size represents the number of bytes transferred.
\end{definition}

\begin{definition}
A quality function determines how efficient the LLVM IR instructions are grouped together in terms of energy consumption, parallelism, load balancing, hardware utilization and inter-cluster data movements.
\end{definition}

The discovery of the processing community structure problem can now be formulated as follows: \textbf{Given} a DDG, \textbf{find} non-overlapping processing communities which \textbf{maximize} the quality function:
\begin{equation}
Q=  \sum_{c=1}^{n_{c}}(\frac{(W_{c} - S_{c})}{W} - \frac{{(W_{c} -  \overline{W} )^2}}{W}) - \frac{\sum_{c=1}^{n_{c}}{E_{N_{c}}}  +  E_{L}}{E}\label{eq}
\end{equation}
and satisfy:
\begin{equation}
N \geq n_{c}\label{eq}
\end{equation}
where $n_{c}$ denotes the number of clusters; $W_{c}$ stands for the sum of edge weights within cluster $c$; $W$ is the sum of all edge weights; $S_{c}$ represents the sum of edges weights connected to cluster $c$; $N$ is the core count.

The first term in equation (7) confines the data flow within the cluster as much as possible. It indicates the difference between the sum of the weights in a cluster and the sum of the weights of the edge connected to the cluster. The greater this term is, the fewer inter-cluster data movements, and the more energy is saved.

The second term in equation (7) measures the standard deviation squared between sum of weights in cluster $c$ and average sum of weights in all clusters. Minimizing this term ensures load balancing and fully takes advantage of parallel execution. 

The third term in equation (7) characterizes the energy model of the application, where $E_{N_{c}}$ calculates the energy consumed at each node using Equation (2) and $E_{L}$ computes the energy consumption for communication transactions. To maximize quality $Q$, this term needs to be minimized in order to save energy.

While optimal communities discovery is an NP-hard problem, we use the Ollivier-Ricci Curvature (ORC) based community detection algorithm \cite{sia2019ollivier} to decide which node should be grouped in which cluster. The complexity of this algorithm is $O(EV^2)$, where $E$ is the number of edges in DDG, and $V$ is the average degree.

\SetKwInput{KwInput}{Input}                
\SetKwInput{KwOutput}{Output}              

\newcommand{\R}{\mathbb{R}}

\begin{algorithm}[h]
\DontPrintSemicolon
  
  \KwInput{Data Dependency Graph (DDG) with a list of nodes and edges and core count $n$}
  \KwOutput{Communities}
 
  $count = 2$\\
  \While{$count \leq n$}
    {
        $DDG' \leftarrow DDG$ Calculate Ollivier-Ricci curvature for all edges;\\
        \While{there is negative edge curve}
        {
        Remove the most negatively curved edge;\\
        Re-calculate the Ollivier-Ricci curvature for the affected existing edges;\\
        }
        PreferentialAttachment($DDG'$, number\textunderscore of\textunderscore communities, minimum\textunderscore community\textunderscore size );
    }
   Pick the solution with the minimum inter-cluster weights;\\

\caption{ORC-based method for Communities Discovery Algorithm}
\end{algorithm}

\subsection{Compact Intelligence Mapping into Constrained Hardware}

The tile to which each cluster is mapped significantly affects the power consumption of the application since it determines the dynamic and static communication cost. Consequently, an approach, which is similar to the one in \cite{hu2003exploiting}, is proposed, but it takes cluster ordering into consideration as well so that it reduces static energy consumption caused by congestion and contention of hardware resources.

\begin{definition}
A task graph (TG) is a weighted directed acyclic graph $TG = G(c_{i},a_{ij},v(a_{ij}),b(a_{ij})|i,j\in N|)$ where each vertex $c_{i}$ represents a cluster of LLVM IR instructions that are grouped together by our community detection algorithm, and each edge $a_{ij}$ represents communication from node $c_{i}$ to node $c_{j}$.
\begin{itemize}
	\item $v(a_{ij})$: data size from $c_{i}$ to $c_{j}$.
	\item $b(a_{ij})$: bandwidth requirement from $c_{i}$ to $c_{j}$.
\end{itemize}
\end{definition}

\begin{definition}
An architecture graph (AG) is a directed graph $AG = G(t_{i},p_{ij},e(p_{ij}))|i,j\in N|)$ where each vertex $t_{i}$ represents a tile, and each edge $p_{ij}$ represents a routing path from $t_{i}$ to $t_{j}$.
\begin{itemize}
	\item $e(p_{ij})$: energy consumption from $t_{i}$ to $t_{j}$.
	\item $L(p_{ij})$: set of links that makes up $p_{ij}$
\end{itemize}
\end{definition}

In order to exploit parallelism and pipelining, we apply topological sort to the task graph before mapping. The depth of cluster $c_{i}$ is defined as the maximum number of edges from the root to $c_{i}$. In Fig. 3, cluster $D$ cannot execute before cluster $B$ and $C$ because it needs data from both of them. However, cluster $B$ and $C$ can execute in parallel because they are at the same depth. 

\begin{figure}[h]
  \centering
  \includegraphics[scale=0.5]{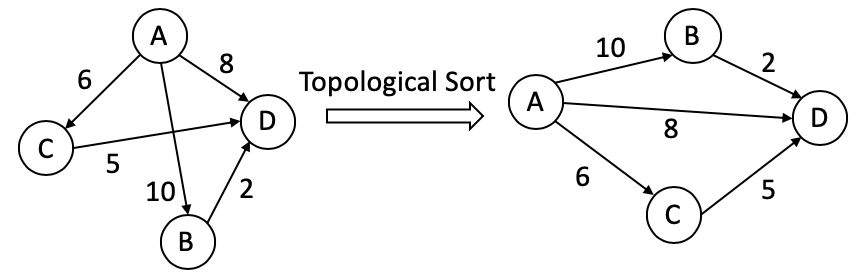}
  \caption{Application of a topological sort to task graph.}
\end{figure}

\SetKwInput{KwInput}{Input}                
\SetKwInput{KwOutput}{Output}              

\begin{algorithm}[h]
\DontPrintSemicolon
  
  \KwInput{$TG$ and $AG$}
  \KwOutput{Mapping from $TG$ to $AG$}
 $count = 0$\\
  \While{$TG$ is not empty}
    {
        \If{$count == 0$}
        {Get the cluster with depth of zero and map to (0,0)}
        \Else
        {Create a set $S_{count}$ of all clusters with depth of $count$;\\
        	Map $S_{count}$ to the available tile in $AG$ so that:\\
        	$min\lbrace E = \sum_{\forall a_{i,j}}^{}{v(a_{i,j})  e(p_{map(c_{i}),map(c_{j})})} \rbrace$}
        	$count++$
    }
   \If{Any idle tile $t$ left in $AG$}
   {
   		Power gate $t$
   }

\caption{Compact Intelligence Mapping Algorithm}
\end{algorithm}

\subsubsection{Energy and Congestion Analysis}
\indent The communication-weighted mapping (CWM) proposed in \cite{hu2003exploiting} (we refer to it as H) fails to consider the order of the clusters, leading to significant potential congestion and static energy consumption in NoCs. This section shows how our communication dependency mapping (CDM) mitigates this problem.

\begin{table*}
  \caption{Mapping comparison: dynamic energy}
  \label{tab:commands}
  \begin{tabular}{|c|c|}
    \toprule
    \multicolumn{2}{|c|}{Dynamic energy = $109 \times 10^{-12}$J}\\
    \midrule
    CWM & CDM\\
    \midrule
    \includegraphics[scale=0.3]{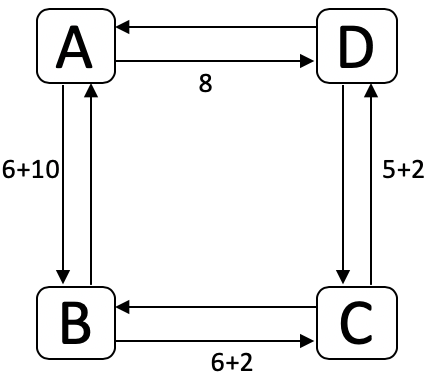} & \includegraphics[scale=0.3]{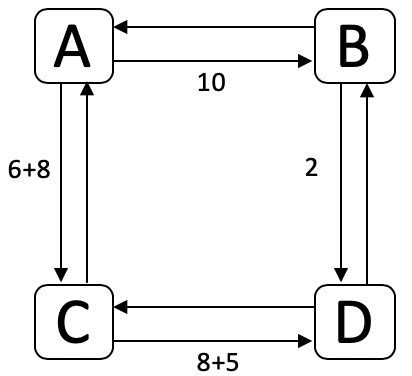}\\
    \bottomrule
  \end{tabular}
\end{table*}

\begin{figure}[h]
  \centering
  \includegraphics[scale=0.5]{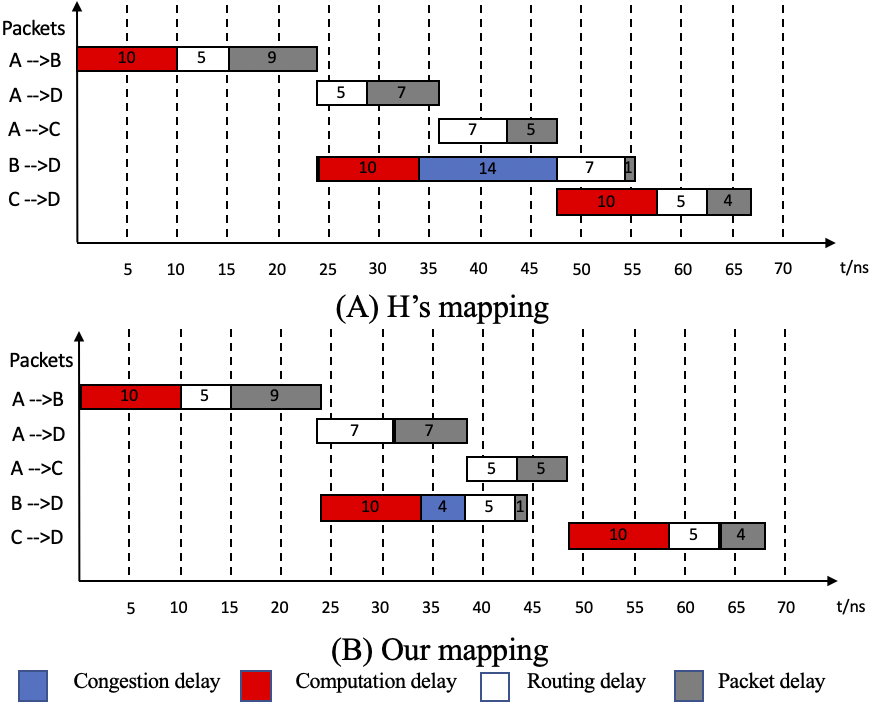}
  \caption{Mapping comparison: static energy.}
\end{figure}

For illustration purposes, we assume $E_{S_{bit}} = E_{L_{bit}} = 1 \times 10^{-12}J/bit$. Applying the CWM to the $TG$ in Fig. 3 may yield the following two different mappings in Table 2. For instance, using Equation (3) in CWM, $E_{Dy_{AC}} = 6 \times (3 \times E_{S_{bit}} + (3-1) \times E_{L_{bit}})= 30 \times 10^{-12}J$. Both mappings' dynamic energy costs are $109 \times 10^{-12} J$.

In terms of static energy, we assume $P_{St} = 1 \times 10^{-12}J$, and the execution time is $10ns$ for all clusters. Also assume one packet flit is $1bit$ and the time for a flit to pass through a switch ($t_{s}$) is $2ns$ and a link ($t_{l}$) is $1ns$. Fig. 4 shows the timing diagram of all computations and all packet deliveries of both mappings. For instance, in CWM,  the first flit of the packet from cluster $A$ to $B$ takes $2\times t_{s} + t_{l} = 5ns$ to arrive (routing delay), while the rest of the packet needs another $9ns$ (packet delay).

In CWM, when cluster $B$ finishes execution and is about to route the packet to $D$, $D$'s input buffer is busy because of $A \rightarrow D$ and $A \rightarrow C$ packet transmissions. Thus, $B$ must wait until $A \rightarrow C$ is done. While the two mappings yield the same execution time of $67ns$, the packets from $B$ to $D$ in CWM experiences a $10ns$ longer congestion delay, hence consuming more static energy. Applying Equations (4) and (5), CWM consumes $17\%$ more energy in interconnect.

\section{Experimental Results}

We use gem5 \cite{binkert2011gem5} together with McPAT \cite{li2009mcpat} for architectural and power simulation. Our hardware model is ARM processor connected in a 2D mesh topology NoC from 4 to 128 cores \cite{agarwal2009garnet} with MESI cache protocol as shown in Table 3. Detailed parameters are listed in TABLE III. We compare our community discovery algorithm with a range of existing scheduling algorithms, including Greedy Scheduling \cite{teodorescu2008variation}, Local Search \cite{winter2008scheduling} and Hierarchical Hungarian \cite{goldberger2008hierarchical}. 

\begin{table*}
  \caption{Parameters of simulation processors}
  \label{tab:commands}
  \begin{tabular}{|c|c|}
    \toprule
    Cores & Up to 128 cores in-order ARM cores at 500MHz\\
    \midrule
    L1 Private Cache & 32KB, 4-way, 32-byte block\\
    \midrule
    L2 Shared Cache & 128KB, 8-way\\
    \midrule
    Topology & 2D Mesh with XY routing\\
    \bottomrule
  \end{tabular}
\end{table*}

First, we examine our ORC-based community discovery algorithm's (Algorithm 1) computational complexity as the number of core grows. The processing community discovery is done offline (only once), so the run time will not affect the controller speed during UAV navigation. Fig. 5 shows the run time overhead of the referenced scheduling against our algorithm over a range of 4 to 128 core organization. For a small number of cores, the run time overhead is acceptable, but increase rapidly for the Local Search and Hierarchical Hungarian, both of which have $O(n^2)$ complexity. Meanwhile, the Greedy Scheduling with $O(n(\log n)^2)$ is more scalable. Their overheads are reasonably small even for 128 cores. 

\begin{figure}[h]
  \centering
  \includegraphics[scale=0.6]{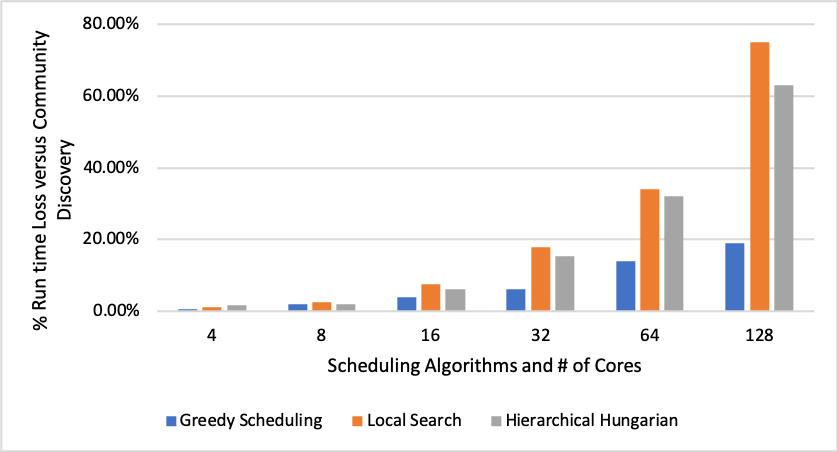}
  \caption{Scheduling runtime overheads over our community discovery algorithm}
\end{figure}

\begin{figure}[h]
  \centering
  \includegraphics[scale=0.6]{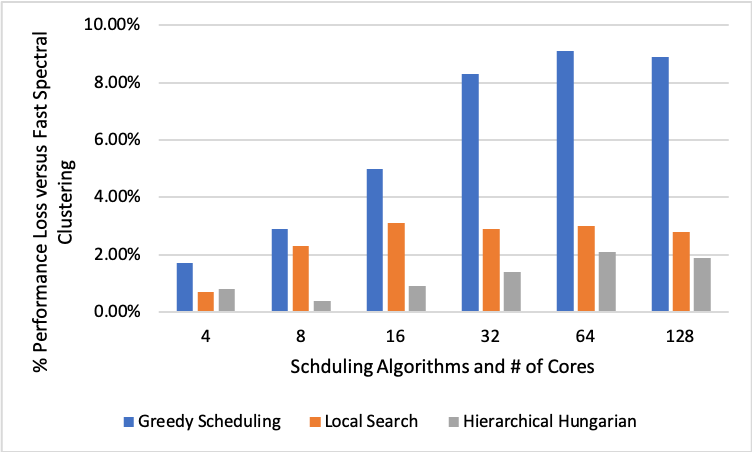}
  \caption{Scheduling performance loss over our community discovery algorithm}
\end{figure}

Fig. 6 evaluates the performance loss of the various scheduling algorithms over our ORC-based communities discovery algorithm. The Greedy Algorithm exhibits the largest performance loss of roughly 9\% for large core count. This is because it fails to consider the complexity of scheduling while only focusing on ranking applications by Instruction Per Clock (IPC). A particular application may suffer significantly if one core has much lower IPC than others. The Local Search delivers decent performance as the core count increases while the Hierarchical Hungarian offers the best combination of run time and performance. However, they both struggle to detect small clusters compared to the graph as a whole because they have a resolution limit. Since ORC-based algorithm takes advantage of the geometric concepts of graph curvature, it identifies better instruction parallelism within the densely-connected community structure and minimizes the communication between clusters. The Data Dependency Graph of UAV application is shown in Fig. 8. 

\begin{figure}[h]
  \centering
  \includegraphics[scale=0.6]{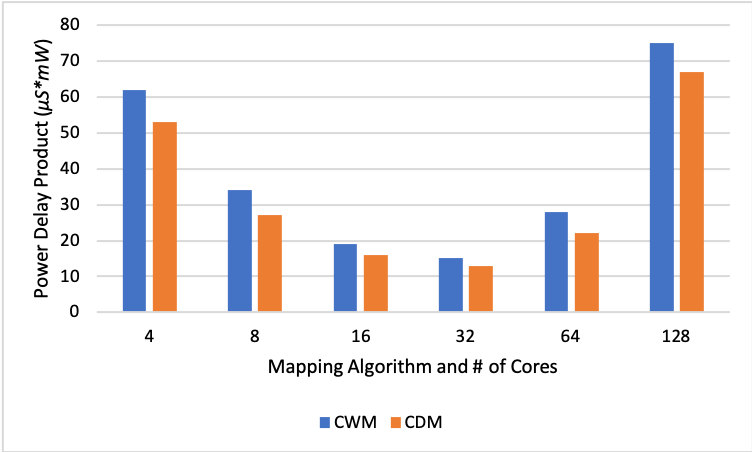}
  \caption{Communication Weighted Mapping versus Communication Dependency Mapping}
\end{figure}

\begin{figure}[h]
  \centering
  \includegraphics[scale=0.2]{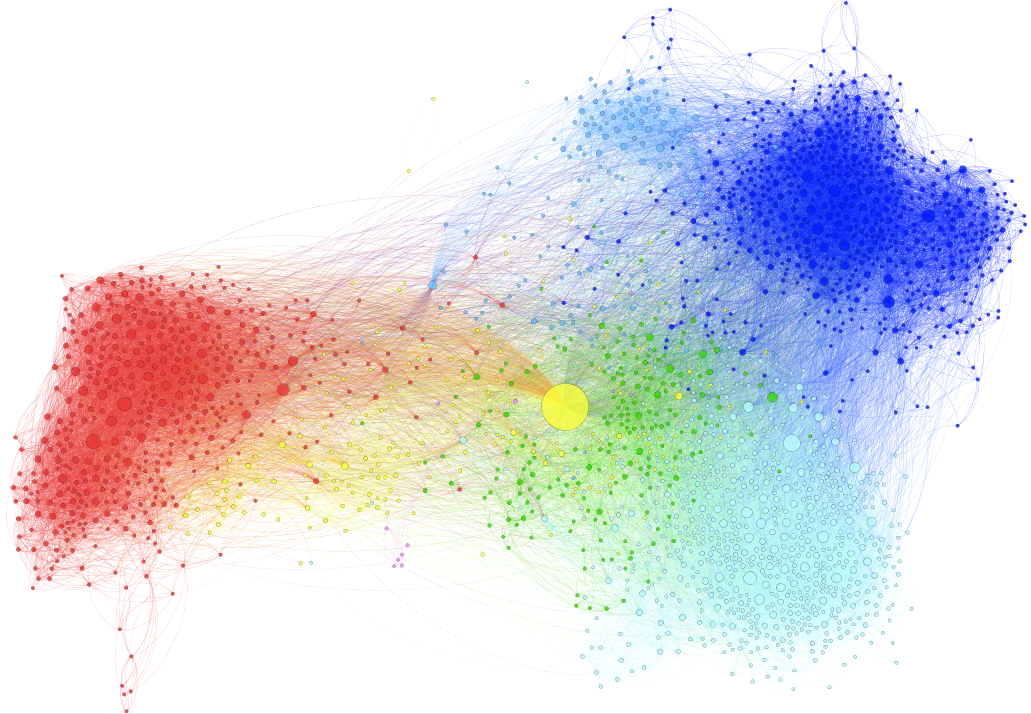}
  \caption{Data Dependency Graph of UAV navigation application.}
\end{figure}

Next, we compare the energy consumption of our CDM and CWM mapping (Fig. 7). Both mappings use the same communities generated by our Fast Spectral Clustering Algorithm. The power values are collected by feeding the outputs from gem5 to McPAT. Having fully taken advantage of parallel execution, load-balancing and optimal inter-community communication, our design has achieved an average of 17\% less Power Delay Product over CWM as the maximum PDP saving of 22\% is achieved for 64-core configuration. As the core count grows, the PDP decreases significantly from 4-core to the lowest at 32-core for 76\%. Starting from 64 cores, the number of flits needed to be routed between cores soars as PDP increases significantly. The lowest PDP is achieved by 32-core configuration at 13.7$\mu S*mW$.

Finally, we illustrate the potential of our design by comparing it with the state-of-art flight controllers used in Intel Aero Ready to Fly Drone \cite{Intel18aero}. As shown in Fig. 9, the Intel Aero Flight Controller is built upon a STM32 MCU, which is a single core ARM Cortex-M4 CPU. We conducted FS simulation in gem5 by running the same navigation algorithm on our 16-core and 32-core configurations and on a single core ARM CPU. As shown in Fig. 10, the navigation algorithm running with our scheduling and mapping scheme has a 3.6x speedup on 16 cores and 4.7x on the 32-core configuration. Meanwhile, the PDP is also improved by 82\% on our 32-core platform.


\begin{figure}[h]
  \centering
  \includegraphics[scale=0.3]{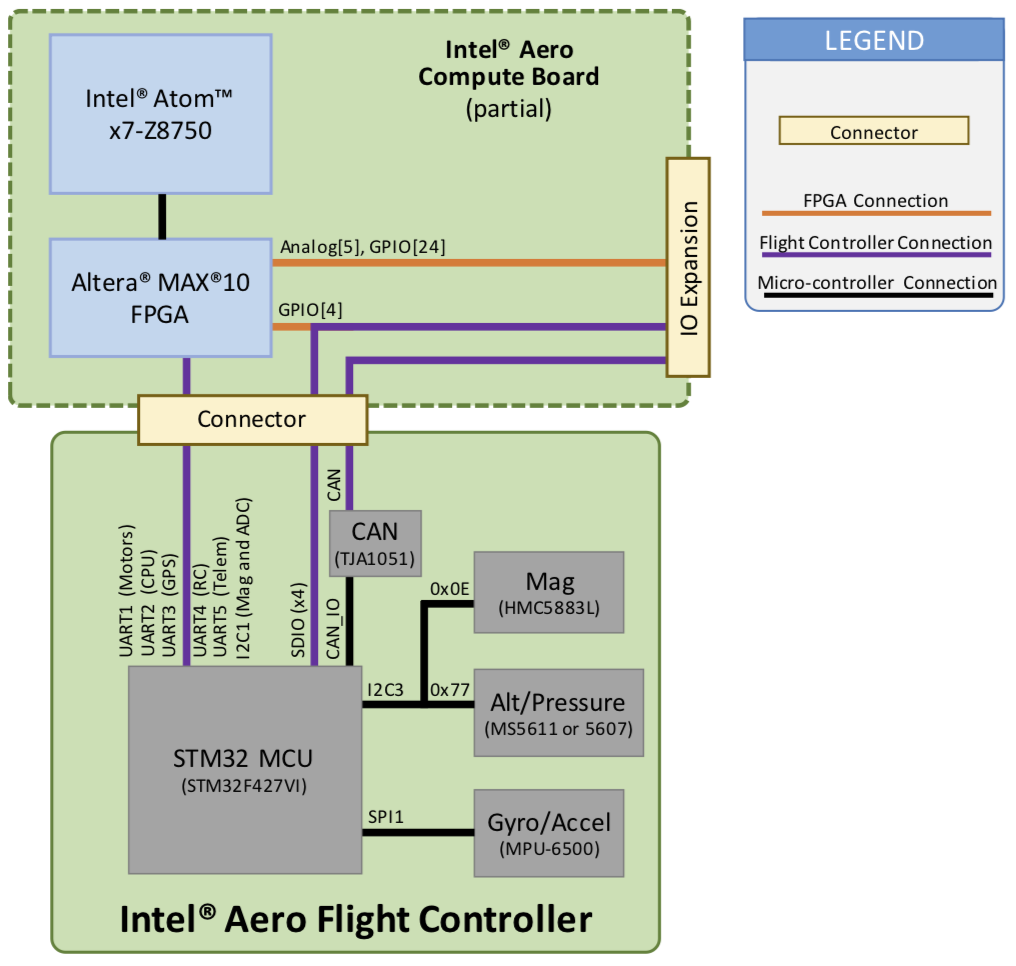}
  \caption{Hardware Block Diagram - Aero Flight Controller}
\end{figure}

\begin{figure}[h]
  \centering
  \includegraphics[scale=0.6]{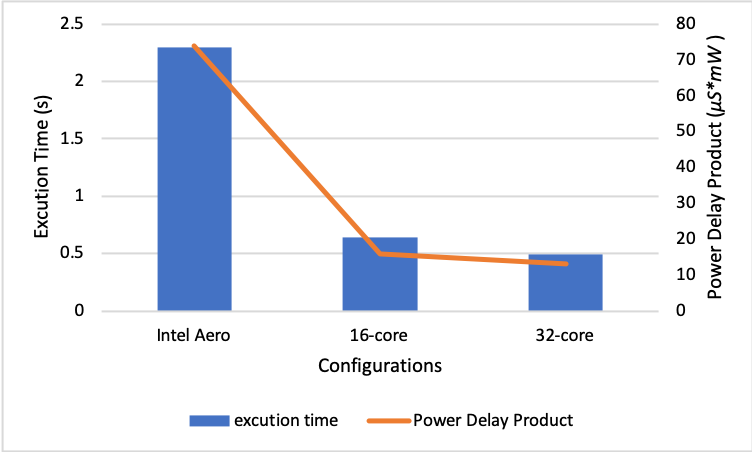}
  \caption{Comparison against Intel Aero Flight Controller}
\end{figure}


\section{Conclusion}

In this paper, we first develop an LLVM IR parser to construct the DDG for UAV autonomous navigation application. Next, we analyze the DDG structure and discover its best parallelization degree by applying our load-balancing and energy-aware processing community discovery algorithm so that data movement is confined within clusters and static energy consumption is minimized. Finally, a congestion-aware mapping scheme based on topological sort is proposed to map clusters onto the NoCs for parallel execution. Simulations show that our optimal 32-core design achieves an average 82\% energy savings and 4.7x performance speedup against the state-of-art Intel Aero Flight Controller.

\bibliographystyle{ACM-Reference-Format}

\begin{thebibliography}{10}

\bibitem{dally2001route}
Dally, William J and Towles, Brian.
\newblock Route packets, not wires: on-chip inteconnection networks.
\newblock In {\em Proceedings of the 38th DAC}, pages 684--689. ACM, 2001.

\bibitem{bogdan2015mathematical}
Bogdan, Paul.
\newblock Power-aware scheduling for AND/OR graphs in real-time systems.
\newblock In {\em Proceedings of the 9th NOCS}, pages 21. ACM, 2015.

\bibitem{bogdan2015mathematical1}
Bogdan, Paul and Xue, Yuankun.
\newblock Mathematical models and control algorithms for dynamic optimization of multicore platforms: A complex dynamics approach.
\newblock In {\em Proceedings of the ICCAD}, pages 170--175. IEEE, 2015.

\bibitem{hu2003exploiting}
Hu, Jingcao and Marculescu, Radu.
\newblock Exploiting the routing flexibility for energy/performance aware mapping of regular NoC architectures.
\newblock In {\em Proceedings of the DATE}, pages 688--693. IEEE, 2003.

\bibitem{xiao2017load}
Xiao, Yao and Xue, Yuankun and Nazarian, Shahin and Bogdan, Paul.
\newblock A load balancing inspired optimization framework for exascale multicore systems: A complex networks approach.
\newblock In {\em Proceedings of the 36th ICCAD}, pages 217--224. IEEE, 2017.

\bibitem{lattner2004llvm}
Lattner, Chris and Adve, Vikram.
\newblock LLVM: A compilation framework for lifelong program analysis \& transformation.
\newblock In {\em Proceedings of the CGO'04}, pages 75. IEEE, 2004.

\bibitem{ye2002analysis}
Ye, Terry Tao and Benini, Luca and De Micheli, Giovanni.
\newblock Analysis of power consumption on switch fabrics in network routers.
\newblock In {\em Proceedings 2002 DAC}, pages 524--529. IEEE, 2002.

\bibitem{grech2015static}
Grech, Neville and Georgiou, Kyriakos and Pallister, James and Kerrison, Steve and Morse, Jeremy and Eder, Kerstin.
\newblock Static analysis of energy consumption for LLVM IR programs.
\newblock In {\em Proceedings of the 18th SCOPES}, pages 12--21. ACM, 2015.

\bibitem{binkert2011gem5}
N.~Binkert, B.~Beckmann, G.~Black, S.~K. Reinhardt, A.~Saidi, A.~Basu, J.~Hestness, D.~R. Hower, T.~Krishna, S.~Sardashti, et~al.
\newblock The gem5 simulator.
\newblock {\em ACM SIGARCH Computer Architecture News}, 39(2):1--7, 2011.

\bibitem{agarwal2009garnet}
N.~Agarwal, T.~Krishna, L.-S. Peh, and N.~K. Jha.
\newblock Garnet: A detailed on-chip network model inside a full-system simulator.
\newblock In {\em 2009 ISPASS}, pages 33--42. IEEE, 2009.

\bibitem{li2009mcpat}
S.~Li, J.~H. Ahn, R.~D. Strong, J.~B. Brockman, D.~M. Tullsen, and N.~P.Jouppi.
\newblock Mcpat: an integrated power, area, and timing modeling framework for multicore and manycore architectures.
\newblock In {\em Proceedings of the 42nd MICRO}, pages 469--480. ACM, 2009.

\bibitem{suleiman2019navion}
A.~Suleiman, Z.~Zhang, L.~Carlone, S.~Karaman, and V.~Sze.
\newblock Navion: A 2-mw fully integrated real-time visual-inertial odometry accelerator for autonomous navigation of nano drones.
\newblock {\em IEEE Journal of Solid-State Circuits}, 2019.

\bibitem{tan2018locus}
C.~Tan, A.~Kulkarni, V.~Venkataramani, M.~Karunaratne, T.~Mitra, and L.-S. Peh.
\newblock Locus: Low-power customizable many-core architecture for wearables.
\newblock {\em TECS}, 17(1):16, 2018.

\bibitem{armah2016feedback}
S.~Armah, S.~Yi, W.~Choi, and D.~Shin.
\newblock Feedback control of quad-rotors with a matlab-based simulator.
\newblock {\em American Journal of Applied Sciences}, 2016.

\bibitem{corke2017flying}
P.~Corke.
\newblock {\em Flying Robots Book: Robotics, Vision and Control}.
\newblock Springer, 2017.

\bibitem{tan2018locus}
Goldberger, Jacob and Tassa, Tamir.
\newblock A hierarchical clustering algorithm based on the Hungarian method.
\newblock {\em Pattern Recognition Letters}, 29(11):1632-1638, 2018.

\bibitem{winter2008scheduling}
Winter, Jonathan A and Albonesi, David H.
\newblock Scheduling algorithms for unpredictably heterogeneous cmp architectures.
\newblock In {\em Proceedings of the 2008 IEEE International Conference on Dependable Systems and Networks With FTCS and DCC (DSN)}, pages 42--51. IEEE, 2008.

\bibitem{teodorescu2008variation}
Teodorescu, Radu and Torrellas, Josep.
\newblock Variation-aware application scheduling and power management for chip multiprocessors.
\newblock {\em ACM SIGARCH computer architecture news}, 36(3):363-374, IEEE Computer Society, 2008.

\bibitem{sia2019ollivier}
Sia, Jayson and Jonckheere, Edmond and Bogdan, Paul.
\newblock Ollivier-ricci curvature-based method to community detection in complex networks.
\newblock {\em Scientific reports}, 9(1):9800, Nature Publishing Group, 2019.

\bibitem{Intel18aero}
Intel. 2018
\newblock Intel Aero Compute Board Hardware Features and Usage.
\newblock Retrieved Oct 27, 2019 from https://www.intel.com/content/dam/support/us/en/documents/drones/development-drones/intel-aero-compute-board-guide.pdf

\bibitem{hoffmann2007quadrotor}
Hoffmann, Gabriel and Huang, Haomiao and Waslander, Steven and Tomlin, Claire.
\newblock Quadrotor helicopter flight dynamics and control: Theory and experiment.
\newblock In {\em AIAA guidance, navigation and control conference and exhibit}, 2017.

\bibitem{rajappa2015modeling}
Rajappa, Sujit and Ryll, Markus and B{\"u}lthoff, Heinrich H and Franchi, Antonio.
\newblock Modeling, control and design optimization for a fully-actuated hexarotor aerial vehicle with tilted propellers.
\newblock In {\em Proceedings of the 2015 IEEE International Conference on Robotics and Automation (ICRA)}, pages 4006--4013. IEEE, 2015.

\bibitem{ullman2017comparing}
Ullman, David G and Homer, Vincent and Horgan, Patrick and Ouellette, Richard. 2017.
\newblock Comparing electric sky taxi visions.

\bibitem{das2009dynamic}
Das, Abhijit and Subbarao, Kamesh and Lewis, Frank.
\newblock Dynamic inversion with zero-dynamics stabilisation for quadrotor control.
\newblock {\em IET control theory \& applications}, 3(3):303--314, IET, 2009.

\bibitem{bouabdallah2004pid}
Bouabdallah, Samir and Noth, Andre and Siegwart, Roland.
\newblock PID vs LQ control techniques applied to an indoor micro quadrotor.
\newblock {\em 22004 IEEE/RSJ International Conference on Intelligent Robots and Systems (IROS)(IEEE Cat. No. 04CH37566)}, 3:2451--2456, IEEE, 2004.

\bibitem{cowling2010direct}
Cowling, Ian D and Yakimenko, Oleg A and Whidborne, James F and Cooke, Alastair K.
\newblock Direct method based control system for an autonomous quadrotor.
\newblock {\em Journal of Intelligent \& Robotic Systems}, 60(2):285--316, Springer, 2010.

\bibitem{lee2009feedback}
Lee, Daewon and Kim, H Jin and Sastry, Shankar.
\newblock Feedback linearization vs. adaptive sliding mode control for a quadrotor helicopter.
\newblock {\em International Journal of control, Automation and systems}, 7(3):419--428, Springer, 2009.

\bibitem{lupashin2010simple}
Lupashin, Sergei and Sch{\"o}llig, Angela and Sherback, Michael and D'Andrea, Raffaello.
\newblock A simple learning strategy for high-speed quadrocopter multi-flips.
\newblock In {\em Proceedings of the 2010 IEEE international conference on robotics and automation}, pages 1642--1648. IEEE, 2010.

\bibitem{lyons2012accelerator}
Lyons, Michael J and Hempstead, Mark and Wei, Gu-Yeon and Brooks, David.
\newblock The accelerator store: A shared memory framework for accelerator-based systems.
\newblock {\em ACM Transactions on Architecture and Code Optimization (TACO)}, 8(4):48, ACM, 2012.

\bibitem{esmaeilzadeh2011dark}
Esmaeilzadeh, Hadi and Blem, Emily and Amant, Renee St and Sankaralingam, Karthikeyan and Burger, Doug.
\newblock Dark silicon and the end of multicore scaling.
\newblock In {\em Proceedings of the 2011 38th Annual international symposium on computer architecture (ISCA)}, pages 365--376. IEEE, 2011.

\end{thebibliography}

\end{document}